\documentclass[aps,pra,reprint,amsmath,amssymb,longbibliography,nofootinbib,floatfix,superscriptaddress]{revtex4-1}
\usepackage{graphicx}
\usepackage{physics}
\usepackage{xcolor}
\usepackage{comment}
\usepackage{makecell}
\usepackage{hyperref}
\usepackage{bm}
\usepackage{letltxmacro}
\usepackage{dutchcal}

\usepackage{mathtools}

\LetLtxMacro{\originaleqref}{\eqref}
\providecommand{i}{\text{i}}

\begin{document}

\title{Vortex formation around islands in random waves}
\author{Alex J. Vernon}
\affiliation{Donostia International Physics Center (DIPC), Donostia-San Sebasti\'an 20018, Spain}

\author{Junyi Ye}
\affiliation{State Key Laboratory of Surface Physics and Institute for Nanoelectronic Devices and Quantum Computing, Fudan University, Songhu Rd., Yangpu Dist., 200438, Shanghai, China}
\affiliation{Key Laboratory of Micro- and Nano-Photonic Structures (Ministry of Education) and Department of Physics, Fudan University, Songhu Rd., Yangpu Dist., 200438, Shanghai, China}

\author{Wenzhe Liu}
\affiliation{State Key Laboratory of Surface Physics and Institute for Nanoelectronic Devices and Quantum Computing, Fudan University, Songhu Rd., Yangpu Dist., 200438, Shanghai, China}
\affiliation{Key Laboratory of Micro- and Nano-Photonic Structures (Ministry of Education) and Department of Physics, Fudan University, Songhu Rd., Yangpu Dist., 200438, Shanghai, China}

\author{Lei Shi}
\affiliation{State Key Laboratory of Surface Physics and Institute for Nanoelectronic Devices and Quantum Computing, Fudan University, Songhu Rd., Yangpu Dist., 200438, Shanghai, China}
\affiliation{Key Laboratory of Micro- and Nano-Photonic Structures (Ministry of Education) and Department of Physics, Fudan University, Songhu Rd., Yangpu Dist., 200438, Shanghai, China}

\author{Konstantin Y. Bliokh}
\affiliation{Donostia International Physics Center (DIPC), Donostia-San Sebasti\'an 20018, Spain}
\affiliation{IKERBASQUE, Basque Foundation for Science, Bilbao 48009, Spain}

\begin{abstract}
Wave vortices are fundamental topological features of interference fields, occurring at nodal points where the wave amplitude vanishes. A distinct class of vortices can instead form around {\it islands} or `holes' in two-dimensional wavefields, where the wave intensity remains finite and may even peak at the boundary. In particular, such vortices occur in M2 ocean tides around New Zealand, Madagascar, Iceland, and Svalbard, yet the conditions governing their appearance have remained elusive. Here we develop a statistical theory of vortices around islands in random two-dimensional wavefields, with and without the Coriolis effect, and test it experimentally. We determine the probabilities of vortices with different topological charges as functions of island size and Coriolis parameter. We find that island-bound vortices emerge with unexpectedly high probability, approaching 50\% in non-rotating systems and nearly 100\% in rotating systems. Moreover, for a broad range of parameters, the presence of a subwavelength island dramatically enhances vortex formation compared with homogeneous random wavefields. Our results explain the formation of tidal vortices around ocean islands of particular sizes ($\sim 0.1$ of the characteristic wavelength) and establish a general mechanism for generating localized high-intensity vortices around defects in diverse wave systems, from water waves to nanophotonic structures.
\end{abstract}

\maketitle

\section{Introduction}

The half-century-long widespread interest in wave vortices began with the recognition of {\it phase singularities} at {\it nodal points} as generic topological features of wavefields by Nye and Berry \cite{Nye1974}. Around such points, the wave phase changes by $2\pi\ell$, where $\ell\in\mathbb{Z}$ is the integer winding number. Vortex waves were subsequently identified as eigenmodes of the orbital angular momentum (OAM) operator \cite{Ceperley1992AJP, Allen1992PRA}, with $\ell$ corresponding to the OAM quantum number. These discoveries laid the foundations of singular optics and optical angular momentum \cite{Soskin2001PO, Dennis2009PO, Allen_book, Andrews_book} and were later extended to acoustics \cite{Hefner1999JASA, Guo2022JAP}, plasmonics \cite{Ohno2006OE, Prinz2023ACSP}, quantum matter waves \cite{Bliokh2017PR, Clark2015Nature, Luski2021Science}, and water waves \cite{Smirnova2024PRL, Wang2025N}. Notably, amphidromic points in {\it ocean tides} constitute one of the earliest observed examples of phase singularities and wave vortices \cite{Nye1988, Berry2001}.

\begin{figure*}[t]
    \centering
\includegraphics[width=0.85\linewidth]{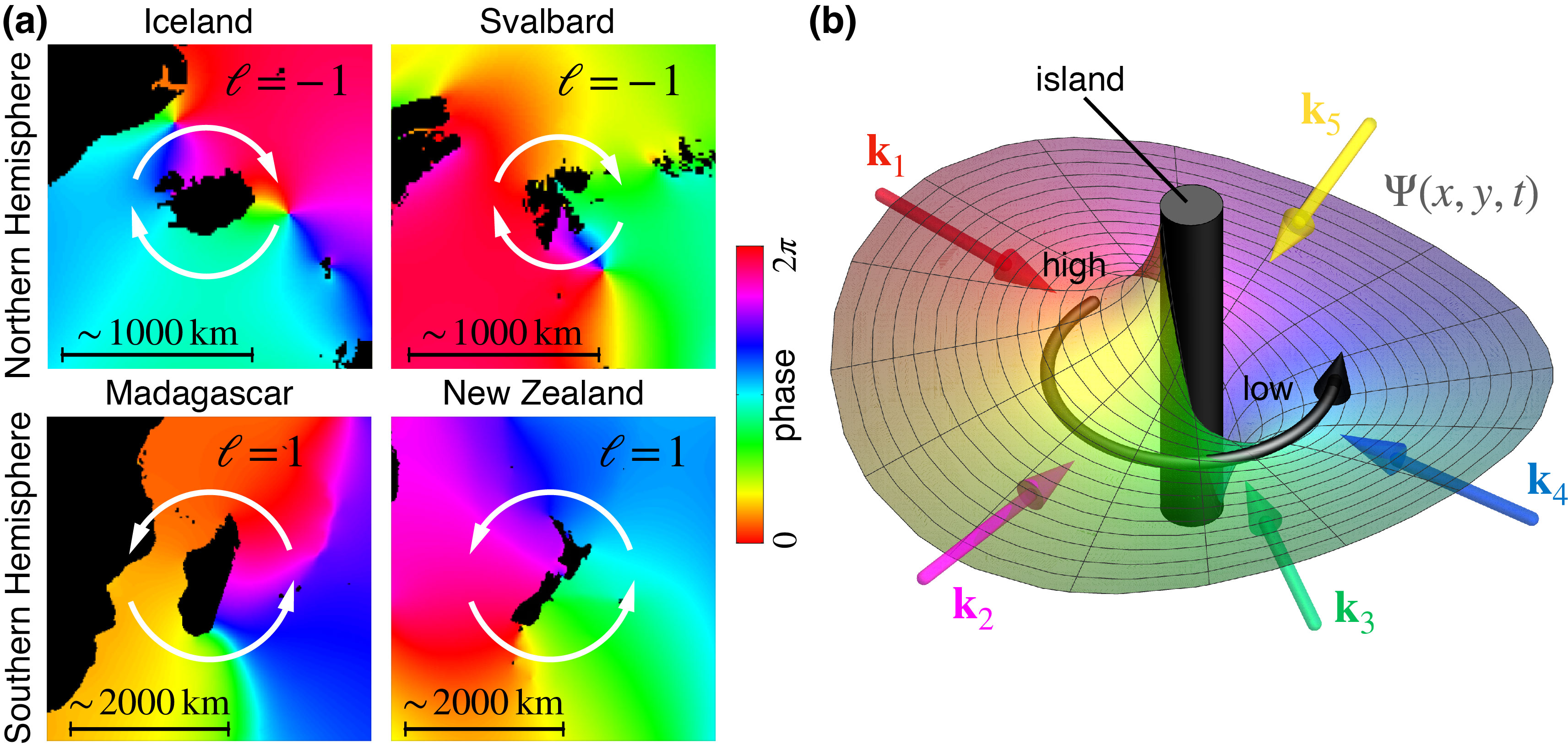}
\caption{{\bf Type-II vortices in tidal waves around ocean islands and schematics of the theoretical model.}  
{\bf (a)} Phase distributions of ocean M2 tides around Iceland and Svalbard in the Northern Hemisphere ($\ell=-1$) and Madagascar and New Zealand in the Southern Hemisphere ($\ell=1$). Corresponding amplitude distributions and examples without vortices ($\ell=0$) are shown in the SM. {\bf (b)} Schematics of the theoretical model. A circular subwavelength island of radius $a$ is illuminated by a random superposition of $M=5$ monochromatic plane waves with frequency $\omega$, wavevectors ${\bf k}_m$ ($m=1,\ldots,5$), random propagation directions, and random phases (indicated by colors). The total field, $\Psi({\bf r},t)=\mathrm{Re}\!\left[\psi({\bf r})e^{-i\omega t}\right]$ (shown by the surface with color-coded phase ${\rm arg}(\psi)$), is the sum of the incident and scattered fields. The field shown here exhibits a type-II vortex with topological charge $\ell=1$: regions of high and low field $\Psi$ on opposite sides of the island revolve around it with angular frequency $\omega$.
}
\label{fig1}
\end{figure*}

Remarkably, tidal waves can also exhibit phase circulation around {\it islands}, rather than nodal points, as observed for semidiurnal (M2) tides around New Zealand \cite{Bye1975, Heath1977, Heath1985, Walters2001, Stevens2021}; see Fig.~\ref{fig1}(a). Such island-bound vortices arise from the nontrivial topology of a {\it multiply connected} 2D space rather than from a phase singularity \cite{Domina2025, Ye2026}, and are also characterized by the topological number $\ell$ quantifying the phase increment around the island. They can also occur around subwavelength holes or defects in plasmonic and polaritonic structures \cite{Gorodetski2010PRB, Vanacore2019NM, Triolo2019SR}.

{Since phase vortices around phase singularities and island-bound vortices (without phase singularities) are topologically distinct, we refer to them as {\it type-I} and {\it type-II} vortices, respectively.}
Unlike conventional type-I vortices, which reside in dark regions around nodal points, type-II vortices are typically accompanied by {\it enhanced wave intensity} at the island boundary and, for subwavelength islands, can concentrate energy and OAM below the diffraction limit \cite{Domina2025}. This qualitative difference originates from the fact that type-I vortices are usually produced by the interference of propagating waves, whereas type-II vortices can involve evanescent near fields localized around the island.

Studies of type-II vortices have so far been limited to a few specific systems: M2 ocean tides around New Zealand \cite{Bye1975, Heath1977, Heath1985, Walters2001, Stevens2021}, plasmonic defects illuminated by normally incident circularly polarized light \cite{Gorodetski2010PRB, Vanacore2019NM, Triolo2019SR}, and the interference of a single incident plane wave with the dipole radiation from an oscillating subwavelength hole \cite{Domina2025, Ye2026}. Consequently, the general conditions for the formation of type-II vortices and their statistical properties in complex wavefields remain unknown. For example, inspection of the global M2 tidal map \cite{Taguchi2014} [see Supplementary Materials (SM)] shows that New Zealand and Madagascar host type-II vortices with $\ell=1$, whereas Iceland and Svalbard exhibit vortices with $\ell=-1$, see Fig.~\ref{fig1}(a). In contrast, Cuba, Novaya Zemlya, and islands smaller than Iceland do not support such vortices (i.e., $\ell=0$) (see SM). What determines whether a given island hosts a vortex, and with what topological charge?

Oceanographic studies have traditionally emphasized the Coriolis effect caused by Earth's rotation, and indeed the vortex handedness in Fig.~\ref{fig1}(a) is perfectly correlated with the sign of the Coriolis parameter in the Northern and Southern Hemispheres. However, the Coriolis effect alone cannot explain the existence of these vortices. It is too weak to trap M2 tidal waves around islands in the manner of coastal Kelvin waves \cite{Longuet-Higgins1967, Longuet-Higgins1968, Longuet-Higgins1969}. Moreover, Madagascar, where the Coriolis effect is relatively weak, hosts a tidal vortex, whereas Novaya Zemlya, at a much higher latitude, does not. Finally, recent experiments and theory have demonstrated that type-II vortices can arise even in non-chiral systems without any Coriolis effect: e.g., around actively oscillating islands \cite{Domina2025, Ye2026}.


In this work, we investigate the {\it statistics} of vortices around islands of different sizes in two-dimensional {\it random} wavefields, both with and without the Coriolis effect. Extending the statistical theory of conventional (type-I) vortices in random wave interference \cite{BerryDennis2000, Hohmann2009PRE, Angelis2016PRL} by including the field {scattered} by the island, we calculate the {\it probabilities} of type-II vortices with different topological charges $\ell$ forming around islands illuminated by a random superposition of incident plane waves, as illustrated in Fig.~\ref{fig1}(b).

Remarkably, we find that neither the Coriolis effect nor active oscillations of the island are required for the formation of type-II vortices. 
Scattering of random incident waves of an island results in surprisingly high probabilities of the type-II vortex occurrence.  
Moreover, the probability of forming a type-II vortex around a subwavelength island of area $S$ can be considerably higher than the probability of finding a conventional type-I vortex within the same area of a homogeneous random wavefield.
We validate predictions of our statistical scattering model by comparison with observed M2 tidal vortices around ocean islands shown in Fig.~\ref{fig1}(a) (where the Coriolis effect is important), and with laboratory water-wave experiments (where the Coriolis effect is negligible), finding excellent agreement in both cases.


Our results explain the origin of tidal vortices around islands and establish general statistical laws for vortex formation around topological defects in multiply connected random wavefields. They also provide a route to engineering high-intensity vortices around subwavelength holes and defects in a broad range of 2D wave systems.

\section{Scattering model}

We consider monochromatic 2D scalar waves (such as the sea-surface elevation of M2 ocean tides) described by the real field $\Psi({\bf r},t)={\rm Re}[\psi({\bf r})e^{-i\omega t}]$, where ${\bf r} = (x,y)$ or $(\rho,\phi)$ in polar coordinates, $\psi({\bf r})$ is the corresponding complex wavefield, and $\omega$ is the wave frequency. We assume that the ${\bf r}$ plane has a circular `hole' or `island' of radius $a$ in the center: ${\bf r} \in \mathbb{R}^2 \backslash D_a$, where $D_a = \{\rho <a \}$. Furthermore, we assume that the waves are affected by the Coriolis effect characterized by the parameter $\Omega$, $|\Omega| < \omega$, obey a linear dispersion relation $\omega = ck_0$ when $\Omega=0$, and are subject to the Neumann-type boundary condition at the boundary of the island. This results in the following wave problem \cite{Longuet-Higgins1969, Bye1975} (see SM):
\begin{equation}
\label{eq_of_motion}
\nabla^2\psi+\frac{\omega^2-\Omega^2}{c^2}\psi=0\,, ~~
\left.\left( \frac{\partial \psi}{\partial\rho} + i\frac{\Omega}{\omega\rho} \frac{\partial \psi}{\partial\phi}\right)\!\right|_{\rho=a} =0\,.  
\end{equation}
Here, the Coriolis effect modifies both the dispersion relation, $\omega^2 -\Omega^2 = c^2 k^2$, and the boundary condition, which enforces a vanishing normal component of the in-plane wave current at the island boundary. 

For M2 tidal waves, $\omega \simeq 2\pi/T_{M2}$ (where $T_{M2} \simeq 44.7 \times 10^3\,$s is the wave period), $c=\sqrt{gh}$, where $g$ is the gravitational acceleration and $h$ is the ocean depth, whereas the Coriolis parameter is given by $\Omega = 2\Omega_E \sin\theta$ \cite{Bye1975}, where $\Omega_{\rm E} = 2\pi/T_E$ is Earth's rotational frequency ($T_E = 86.4 \times 10^3\,$s) and $\theta$ is the island latitude.

The problem \eqref{eq_of_motion} has previously been studied mainly in the strong-Coriolis regime, $|\Omega| > \omega$, where it supports trapped Kelvin waves around islands \cite{Longuet-Higgins1969}. Here we instead focus on the weak-Coriolis regime, $|\Omega| < \omega$, and subwalength islands, $k_0a < 2\pi$, demonstrating that this parameter range supports robust type-II vortices even in the absence of the Coriolis effect ($\Omega=0$).

Note that the model \eqref{eq_of_motion} is not restricted to ocean tides and applies generally to 2D wave systems containing holes or defects, including plasmonic, metamaterial, or graphene waves, where a Coriolis-like term can be induced by an external or synthetic magnetic field. The Neumann boundary condition is essential though: subwavelength holes with Dirichlet boundary conditions have recently been shown to produce only negligible perturbations of generic wavefields \cite{Berry2024}. We also emphasize that Eqs.~\eqref{eq_of_motion} constitute an idealized model of tidal waves around real islands, neglecting irregular coastlines, continental shelves, and spatial variations of the ocean depth.


To analyze the appearance of type-II vortices around the island, we solve the {\it scattering} problem defined by Eqs.~\eqref{eq_of_motion}. 
The incident field is taken as a superposition of $M$ plane-wave solutions of Eqs.~\eqref{eq_of_motion} propagating in azimuthal directions $\phi_m$ and having phases $\alpha_m$, $m=1,...,M$.
Due to the circular geometry, the scattered field is naturally expanded into a series of outgoing cylindrical waves, $H_n^{(1)}\!(k\rho)e^{in\phi}$, where $H_n^{(1)}$ are Hankel functions of the first kind. This yields the total (incident plus scattered) wavefield:
\begin{equation}
\label{total_field}
\psi=\underbracket{\sum_{m=1}^{M}e^{i [k\rho\cos(\phi-\phi_m)+\alpha_m]}}_{\psi_\text{inc}}+\underbracket{\sum_{n=-\infty}^\infty c_nb_nH_n^{(1)}\!(k\rho)e^{i n\phi}}_{\psi_\text{sc}}.
\end{equation}
Here, the overall constant amplitude is omitted, the coefficients $c_n=\sum_{m=1}^{N}e^{i(\alpha_m-n\phi_m)}$ are determined by the phases and directions of the incident plane waves, whereas the scattering coefficients are obtained from the boundary condition in Eq.~\eqref{eq_of_motion} (see SM):
\begin{equation}
\label{coriolis_coeff}
b_n=-i^n\frac{n\chi J_n(ka)-kaJ'_{n}(ka)}{n\chi H^{(1)}_n\!(ka)-kaH^{(1)\prime}_{n}\!(ka)}\,,
\end{equation}
where $J_n$ are Bessel functions of the first kind, the prime denotes derivative with respect to the argument, and $\chi = \Omega/\omega \in [-1,1]$ is the dimensionless Coriolis parameter.

The presence of a type-II vortex is characterized by the topological winding number of the phase of the total wavefield around the island: $\ell =\dfrac{1}{2\pi}\int\limits_0^{2\pi} {\rm Im}\!\left[\dfrac{\partial\ln\psi (a,\phi)}{\partial\phi}\right]d\phi$.
A vortex with $\ell \neq 0$ requires broken mirror symmetry $\phi \to -\phi$. In the absence of the Coriolis effect, $\chi=0$, this asymmetry can be provided by the interference of multiple incident plane waves through the coefficients $c_n$ for positive and negative $n$. When the Coriolis effect is present, $\chi \neq 0$, the vortex effect can be enhanced by the asymmetry of the coefficients $b_n$. 

\begin{figure}[t]
\centering
\includegraphics[width=0.9\columnwidth]{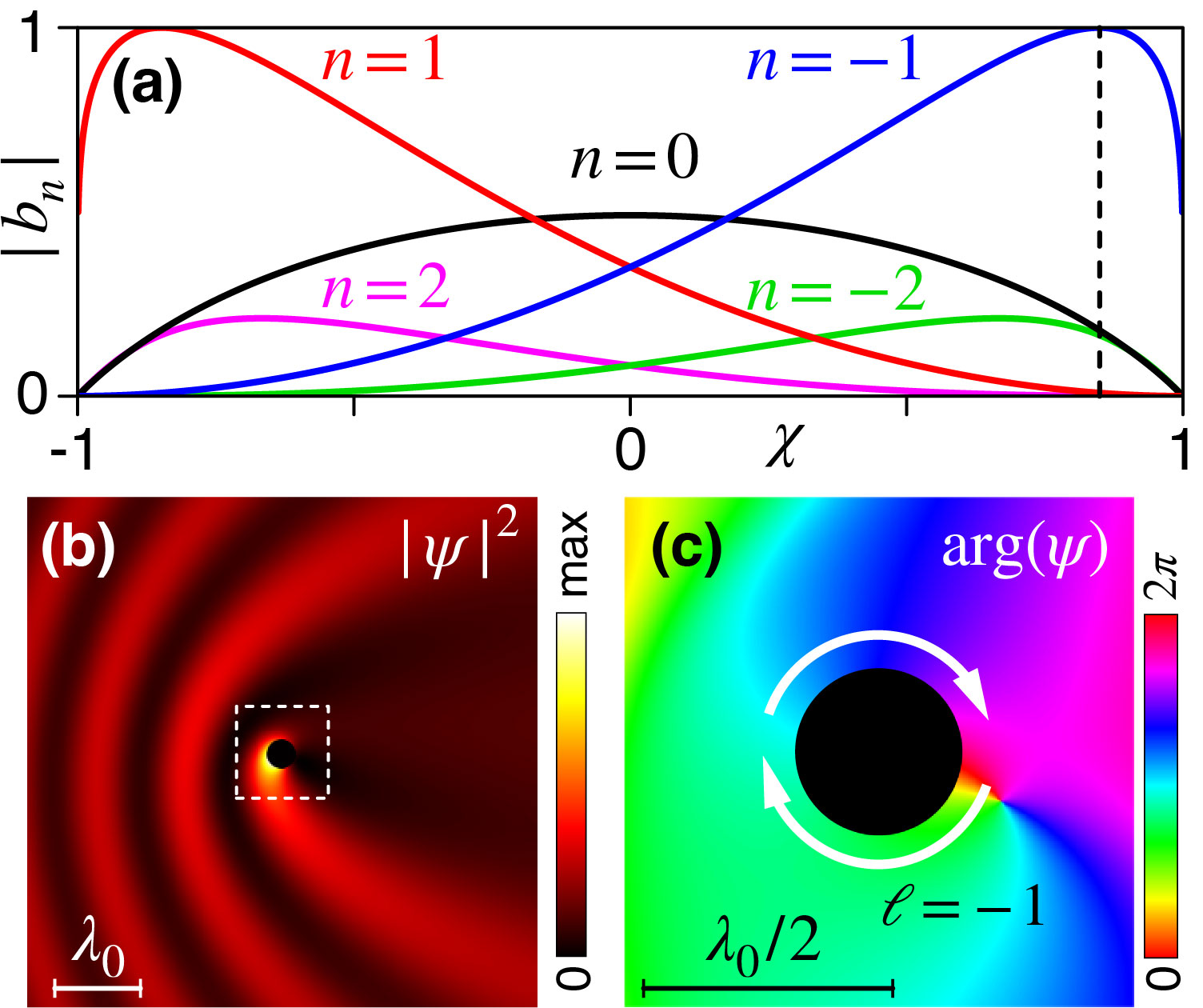}
\caption{{\bf Excitation of a type-II vortex by a single incident plane wave in the presence of the Coriolis effect.}
{\bf (a)} Magnitudes of the scattering coefficients, $|b_n |$, Eq.~\eqref{coriolis_coeff}, versus the Coriolis parameter $\chi$ for a subwavelength island of radius $a=1/k_0$ and different orders $n=-2,...,2$.
{\bf (b)} Intensity distribution, $|\psi|^2$, of the approximate field $\psi({\bf r}) \simeq e^{i k\rho\cos\phi}-i H_1^{(1)}(k\rho)e^{-i\phi}$ for $\chi\simeq 0.85$ [dashed line in {\bf (a)}].
{\bf (c)} Corresponding phase distribution, ${\rm arg}(\psi)$, zoomed in from the dashed region in {\bf (b)}, exhibiting a type-II vortex of charge $\ell=-1$ around the island and a nearby type-I vortex of opposite charge. The scalebars use $\lambda_0 = 2\pi/k_0$.}
\label{fig2}
\end{figure}

Figure~\ref{fig2} illustrates this mechanism for a single incident plane wave, $M=1$, with $\phi_1=\alpha_1=0$, which impinges on a subwavelength island of radius $a=1/k_0$ (where $k_0 = \omega/c$).
The magnitudes of the lower-order scattering coefficients $b_n$, $n=-2,...2$ (higher-order coefficients are negligible) are plotted in Fig.~\ref{fig2}(a) as functions of the Coriolis parameter $\chi$. One can see that positive values of $\chi$ favor negative-order cylindrical harmonics, and vice versa.
Choosing $\chi\simeq0.85$, the coefficient $|b_{-1}| \simeq 1$ dominates the scattered field, so that $\psi\simeq e^{i k\rho\cos\phi}-i H_1^{(1)}(k\rho)e^{-i\phi}$, representing the interference of the incident plane wave with a high-intensity wave vortex with topological charge $\ell=-1$.
The intensity and phase distributions of this approximate total field, shown in Fig.~\ref{fig2}(b,c), reveal a type-II vortex with $\ell=-1$, circulating around the island boundary, together with a nearby type-I vortex with $\ell=1$.
Importantly, the type-II vortex is accompanied by a strong intensity enhancement over most of the island boundary.

\section{Statistics of type-II vortices: theory and ocean tides}
\label{sec:statistics}

Using the model \eqref{eq_of_motion}--\eqref{coriolis_coeff}, we now address the central question of this work: what is the {\it probability} of a type-II vortex with topological charge $\ell$ emerging around an island of radius $a$, with the Coriolis parameter $\chi$, in a {\it random} wavefield? To this end, we consider an incident field comprising $M=5$ plane waves with random azimuthal directions $\phi_m$ and phases $\alpha_m$, independently and uniformly distributed over $[0,2\pi)$ (we have verified that choosing larger $M$ does not produce a significant difference in the vortex statistics). 
Note that in the absence of the island and scattered field, this incident field produces a random interference wavefield underlying the statistics of conventional type-I vortices \cite{BerryDennis2000, Hohmann2009PRE, Angelis2016PRL}, see Section~\ref{sec:Experiment}.

\begin{figure*}[t]
\centering
\includegraphics[width=0.9\linewidth]{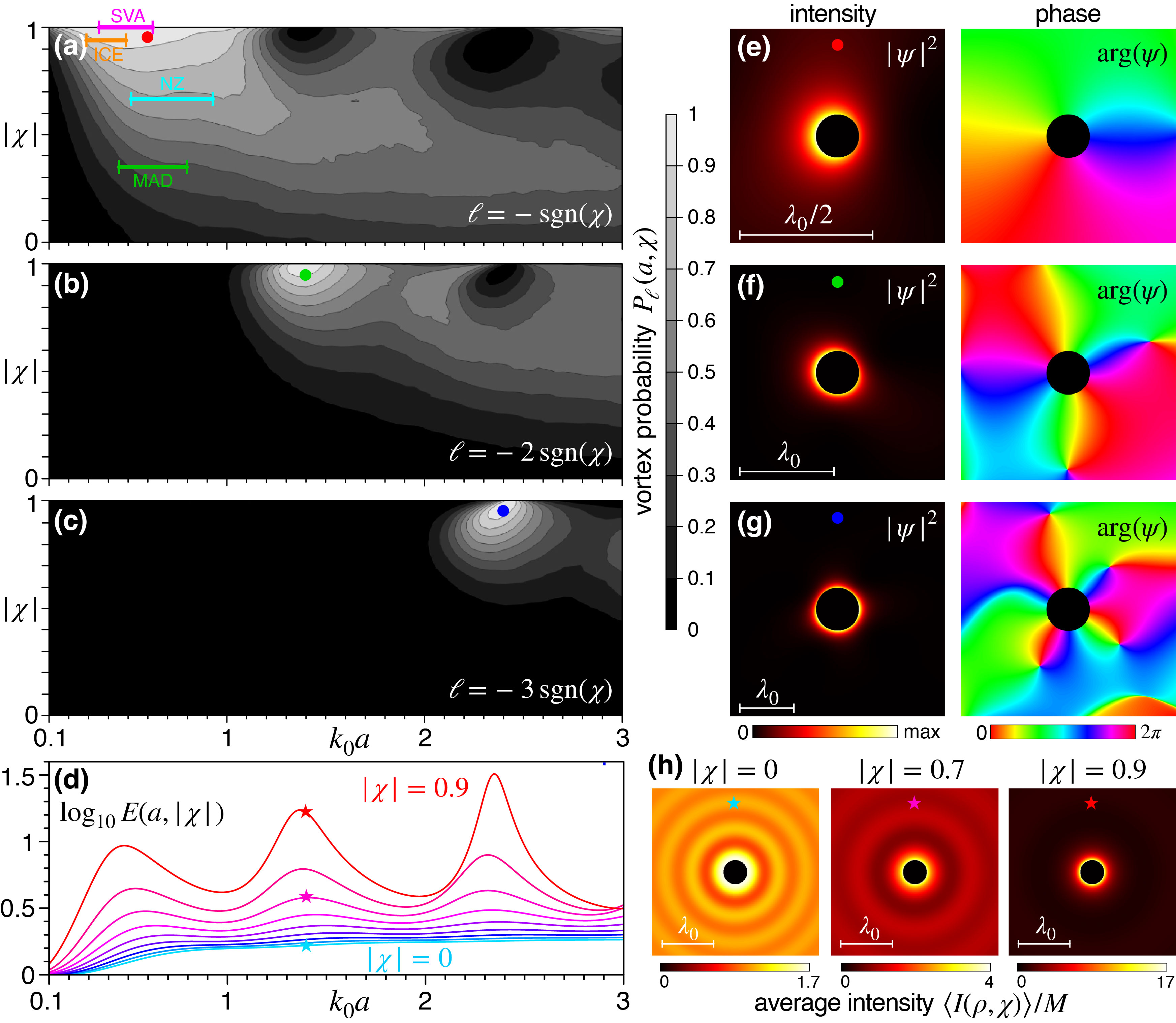}
\caption{{\bf Statistics of type-II vortices around islands in random wavefields.} 
{\bf (a)-(c)} Numerically-calculated probabilities of emerging type-II vortices with topological charges $\ell=-{\rm sgn}(\chi)$, $-2\,{\rm sgn}(\chi)$, and $-3\,{\rm sgn}(\chi)$ as functions of the dimensionless island radius $k_0a$ and the Coriolis parameter $|\chi| \in [0,1]$. 
The estimated parameters of Iceland (ICE), Svalbard (SVA), Madagascar (MAD), and New Zealand (NZ) [Fig.~\ref{fig1}(a)], listed in Table~\ref{tab:islands}, are overlaid in the form of $\pm 35\%$ error bars reflecting large uncertainty in estimations of $k_0 a$.
{\bf (d)} Logarithmic plots of the average-intensity enhancement $E (a,|\chi|) = \langle I(a,|\chi|) \rangle/M$ at the island boundary versus $k_0 a$ for different values of $|\chi| \in [0,0.9]$.
{\bf (e)-(g)} Intensity (left column) and phase (right column) distributions of representative examples of randomly generated wavefields for the $(k_0a,|\chi|)$ parameters corresponding to resonant quasi-trapped modes, indicated by the red, green, and blue dots in {\bf (a)-(c)}.
{\bf (h)} Ensemble-averaged intensity distributions $\langle I(\rho,|\chi|) \rangle$ for $k_0a = 1.4$ and different values of $|\chi|$, corresponding to the cyan, magenta, and red stars in {\bf (d)} and reflecting the emergence of the quasi-trapped mode with $\ell = -2\,  {\rm sgn(\chi)}$ as $|\chi|$ approaches 1. The scalebars use $\lambda_0 = 2\pi/k_0$. See also SM for the non-vortex probability $P_0(a,\chi)$ and average intensity distributions at different values of $k_0 a$ and $|\chi|$.
}
\label{fig3}
\end{figure*}

For each parameter pair $(a,\chi)$, we solve the scattering problem \eqref{eq_of_motion}--\eqref{coriolis_coeff} numerically for 1000 independent realizations of the incident field and determine the probabilities $P_\ell(a,\chi)$ of obtaining type-II vortices with $\ell=0,\pm1,\pm2,\pm3$.
In doing so, we restrict the scattered field by the orders $|n|\le5$ (we have verified that higher-order harmonics have negligible amplitudes and do not affect the statistics).
Figures~\ref{fig3}(a)-(c) show the probabilities of the appearance of vortices with $\ell=-{\rm sgn}(\chi)$, $-2\,{\rm sgn}(\chi)$, and $-3\,{\rm sgn}(\chi)$ as functions of the island radius $a$ and the Coriolis strength $|\chi|$. (See also SM for the non-vortex $\ell=0$ probability $P_0(a,\chi)$; vortices with ${\rm sgn}(\ell) = {\rm sgn}(\chi)$ occur with lower probabilities.) Several remarkable features can be seen in these plots. 


First, in the absence of the Coriolis effect, $\chi=0$, the probability $P_{1}(a,0)$ increases from zero at $a=0$ and approaches $\simeq 0.25$ for $k_0a \gtrsim 1$. 
Since vortices with $\ell=\pm1$ are equally probable in this case, the total probability of forming a $|\ell|=1$-charged vortex, $P_{|\ell|=1}=P_{-1}+P_1$, approaches $\simeq 0.5$ (see SM and Section~\ref{sec:Experiment}).
Thus, type-II vortices around subwavelength islands are not rare, even without the Coriolis effect and without active island oscillations considered in \cite{Domina2025, Ye2026}.


Second, the Coriolis effect, $0<|\chi|<1$, can strongly enhance the formation of vortices with ${\rm sgn}(\ell) = - {\rm sgn}(\chi)$. 
As $|\chi|$ approaches unity, the probabilities $P_\ell(a,\chi)$ develop pronounced resonant peaks, reaching $P_\ell \simeq1$ at $k_0a_{\rm res} \simeq 0.4$, $1.4$, and $2.4$, for  $\ell=-{\rm sgn}(\chi)$, $-2\,{\rm sgn}(\chi)$, and $-3\,{\rm sgn}(\chi)$, respectively. These resonances coincide approximately with the maxima of the corresponding scattering coefficients, $|b_\ell|=1$, obtained from Eq.~\eqref{coriolis_coeff} when $\ell\chi Y_\ell(ka_{\rm res}) -kaY'_\ell(ka_{\rm res})=0$, where $ Y_\ell$ are Bessel functions of the second kind (see SM).
Although true trapped modes exist only for $|\chi|>1$, where the wavenumber $k$ becomes imaginary \cite{Longuet-Higgins1969}, sufficiently large values of $|\chi|<1$ support {\it quasi-trapped modes}. Near resonance, the wavefield is well approximated by the corresponding cylindrical harmonic, $\psi_{\rm res} \propto H_\ell^{(1)}\!(k\rho) e^{i\ell\phi}$, see Fig.~\ref{fig3}(e)--(g). 


A distinctive feature of type-II vortices is the strong field {\it enhancement} at the boundary of a subwavelength island \cite{Domina2025, Ye2026}. Consequently, their presence in a random wavefield is reflected in the {\it ensemble-averaged intensity}. Averaging the intensity, $I=|\psi|^2$, of the field \eqref{total_field} over the random propagation directions $\phi_m$ and phases $\alpha_m$ yields (see SM) 
$\langle I (\rho,|\chi|) \rangle = M\! \left[ 1 + 2 \Re\sum_ni^{-n}b_nJ_n(k\rho)H_n(k\rho) + \sum_n |b_n|^2 |H_n(k\rho)|^2 \right]$.
The corresponding intensity enhancement at the island boundary (relative to the average incident-field intensity $M$),
$E (a,|\chi|) = \langle I(a,|\chi|) \rangle/M$,
is shown in Fig.~\ref{fig3}(d).
Even without the Coriolis effect, the enhancement is noticeable, reaching $E(a,0) \gtrsim 1.5$ for $k_0a \gtrsim 1$. As $|\chi|\to 1$, the wavenumber tends to zero, $k\to 0$, and the enhancement diverges, $E(a,1)\to \infty$. For resonances corresponding to quasi-trapped modes at $|\chi|=0.9$, the enhancement reaches $E(a_{\rm res},0.9)\simeq10$--30. The average intensity is well approximated by $\langle I_{\rm res} \rangle  \simeq M |b_\ell|^2 |H_\ell^{(1)}\!(k\rho)|^2$ in this case (see SM).


\begin{table}[t]
\caption{{\bf Approximate parameters of the ocean islands considered in this work.}}
\begin{tabular}{c c c c c}
\hline \hline 
{Island}  & Coriolis, $\chi$ & Radius, $a$ & Depth, $h$ & $k_0 a$ \\ [0.5em]  \hline
{\bf Iceland}  
  & $0.94$  & $\sim 270$~km & $\sim 1$~km & $\sim 0.38$ \\ [0.5em]
{\bf Svalbard} 
  & $1$  & $\sim 240$~km & $\sim 0.5$~km & $\sim 0.48$ \\ [0.5em]
{\bf Madagascar}        
  & $-0.34$  & $\sim 770$~km & $\sim 3$~km & $\sim 0.63$ \\ [0.5em]
{\bf New Zealand}      
  & $-0.68$  & $\sim 730$~km & $\sim 2$~km & $\sim 0.73$  \\  [0.5em]
\hline \hline
\end{tabular}
\label{tab:islands}
\end{table}

We now compare these predictions with the observed type-II vortices in the M2 tides around Iceland, Svalbard, Madagascar, and New Zealand [Fig.~\ref{fig1}(a)]. For these islands, the dimensionless Coriolis parameter is 
$\chi = (2\Omega_E/\omega) \sin \theta = (2 T_{M2}/T_E) \sin \theta \simeq 1.035 \sin \theta$,
and the corresponding values are listed in Table~\ref{tab:islands}.
Estimating the second parameter, $k_0 a$, is less straightforward because our model assumes a circular island in an ocean of uniform depth, whereas real islands have irregular coastlines, are surrounded by highly inhomogeneous bathymetry, and often sit on continental shelves that further modify tidal-wave propagation \cite{Bye1975}.
We therefore estimate the characteristic island radius $a$ as that of the circumscribed circle enclosing the island (without its shelf), and estimate the representative ocean depth $h$ along this circle from global bathymetric maps. The obtained dimensionless size parameters, 
$k_0 a =  \omega a/\sqrt{gh}$, are listed in Table~\ref{tab:islands}.
Finally, the resulting `coordinates' of each island in the $(|\chi|,k_0a)$ plane are plotted in Fig.~\ref{fig3}(a) with $\pm35\%$ error bars in $k_0a$ indicating the uncertainty of these estimates.

Despite these simplifications, the agreement between theory and observations is striking. All four islands fall within the region of enhanced probability for vortices with $\ell=-\mathrm{sgn}(\chi)$, with typical radii $a\sim 0.5/k_0 \sim 0.1 \lambda_0$, demonstrating that our statistical model captures the essential physical mechanism underlying the formation of tidal vortices around large, yet still subwavelength, ocean islands.
These results suggest that tidal vortices around islands are manifestations of a universal wave-scattering mechanism that extends far beyond oceanography.

\begin{figure*}[t!]
\centering
\includegraphics[width=0.9\linewidth]{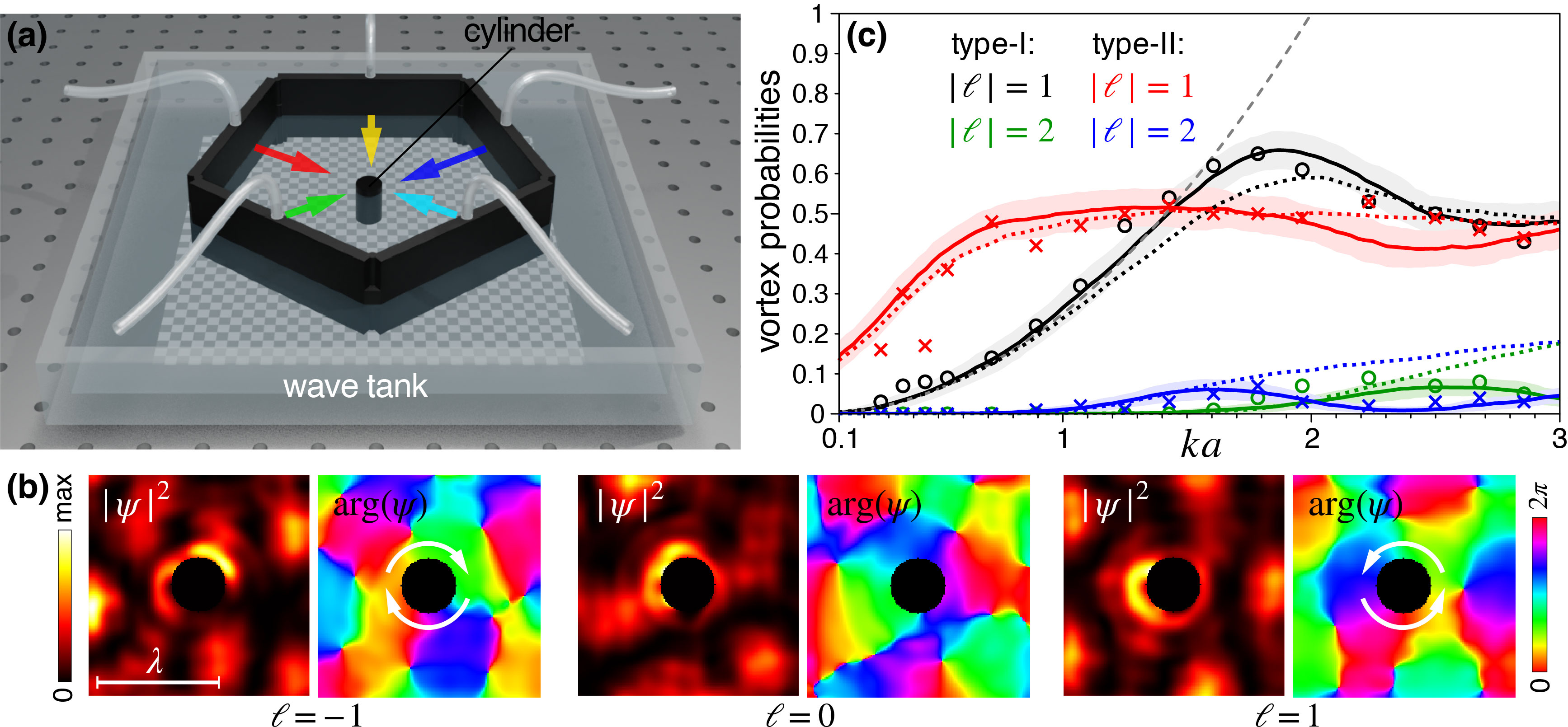}
\caption{{\bf Laboratory measurements of the statistics of type-II and type-I vortices.} {\bf (a)} Schematics of the gravity-capillary-wave experiment, which involves $M=5$ incident plane waves with fixed directions but random phases and amplitudes. The cylinder of radius $a$ in the center plays the role of island.
{\bf (b)} Examples of measured water-wave fields with topological charges $\ell=-1$, $0$, and $1$ around a cylinder with $ka \simeq 0.7$. 
{\bf (c)} Measured (symbols) and numerically calculated (curves) probabilities of vortices with different topological charges. Type-I vortices (phase singularities) were measured in the absence of the cylinder, whereas type-II vortices were measured for cylinders of different radii $a$.
Shaded bands indicate one standard deviation of the numerical results.
The dashed curve shows the theoretical probability of a singly charged type-I vortex for $ka\ll 1$, based on the phase-singularity density derived in~\cite{BerryDennis2000}: $k^2a^2/4$. The dotted curves show numerical results for $M=50$ incident plane waves with random phases, uniformly-distributed propagation directions, and equal amplitudes, which approximately correspond to ``isotropic random waves'' considered in~\cite{BerryDennis2000}.
}
\label{fig4}
\end{figure*}

\section{Laboratory observation and comparison with type-I vortices}
\label{sec:Experiment}

We now focus on the non-Coriolis case, $\chi=0$, $k=k_0$. The purpose is two-fold: (i) to perform a laboratory measurement of the statistics of type-II vortices in a setup closely matching our theoretical model; and (ii) to compare the statistics of type-II vortices with the previously-studied statistics of type-I vortices (i.e., phase singularities) in random wavefields \cite{BerryDennis2000, Hohmann2009PRE, Angelis2016PRL}. 

The experiment was performed using the setup shown in Fig.~\ref{fig4}(a) and involving a $50 \times 50~\mathrm{cm}^2$ wave tank with depth $h = 1.5$~cm. The driving frequency was $\omega/(2\pi) = 7.2$~Hz, corresponding to a gravity-capillary wavelength of $\lambda = 2\pi/k \simeq 3.5$~cm. The incident field, consisting of $M = 5$ plane waves, was generated by a pentagonal structure with 20-cm sides, each containing a 18-cm wave-emitting slot and driven by speakers controlled through a multichannel sound card (Orion 32+, Antelope Audio), externally interfaced with a computer. 
Since it was difficult to vary the propagation directions of incident plane waves (they were fixed as $\phi_m \simeq 2\pi m/M$), we randomly varied the phases $\alpha_m \in [0,2\pi)$ and the dimensionless amplitudes {$A_m \in [0.4,1.6]$} of the incident waves {(to be included under the sum in the first term in Eq.~\eqref{total_field} and in the scattering coefficients $c_n$)}. 
We used MATLAB's ``rand" function to generate uniformly distributed random values within the specified ranges for phases and amplitudes. 
{The same set of 100 random-field realizations was used for each set of measurements.}

Cylinders of different radii, representing circular islands, were placed in the center of the tank for different sets of measurements. The resulting wavefields $\Psi({\bf r},t)$ were measured using the top-view camera and Fast Checkerboard Demodulation (FCD) method \cite{Wildeman2018, Wang2025N}. 
At each spatial point ${\bf r}$, a Fast Fourier Transform (FFT) was performed on the temporal signal, followed by the application
of a Gaussian frequency window centered at 7.2~Hz with a $1/e$  amplitude half-width of 1~Hz. The corresponding complex field $\psi({\bf r})$ was then reconstructed via the Hilbert transform. 
Examples of the measured fields with topological charges $\ell=-1$, $0$, and $1$ for a cylinder of radius $a=0.4$~cm ($ka\simeq0.7$) are shown in Fig.~\ref{fig4}(b).

Note that the wavefield extracted using the FCD method may contain spurious signals near the cylinder boundary. Therefore, following the procedure of \cite{Wildeman2018}, we applied a slightly enlarged mask with Gaussian-smoothed edges prior to FCD processing and excluded a central circular region whose radius was $a+0.1\lambda$.

In the first set of experiments, the cylinder was removed and the random interference field, $\psi_{\rm inc}({\bf r})$, was measured. 
The topological charge $\ell$ was then evaluated along circles of different subwavelength radii $a$ centered in the field.
The fraction of realizations with $|\ell|=1$ approximately corresponds to the probability that a conventional type-I vortex (phase singularity) is enclosed by the circle.
The measured probabilities, as well as numerical simulations, agree closely with the theoretical prediction obtained from the density of phase singularities in two-dimensional random wavefields, $d=k^2/(4\pi)$ \cite{BerryDennis2000}, multiplied by the circle area, $\pi a^2$, for $ka\lesssim 1.5$; see Fig.~\ref{fig4}(c).
For larger radii, the probability of enclosing two or more phase singularities becomes non-negligible. In particular, the $|\ell|=2$ curve for type-I vortices in Fig.~\ref{fig4}(c) corresponds predominantly to two enclosed vortices of the same sign, while the $|\ell|=1$ probability consequently deviates from the simple $\pi a^2d$ behavior.


In the second set of experiments, cylinders of different radii $a$ were placed at the center of the tank, and the total (incident plus scattered) wavefield was measured. The topological charge was evaluated around the masked boundary at radius $a+0.1\lambda$, and the probabilities of type-II vortices with $|\ell|=1$ and $2$ were determined. The measured probabilities agree very well with the predictions of our model, see Fig.~\ref{fig4}(c) and SM. In particular, the probability of forming a singly charged type-II vortex reaches $P_{|\ell|=1}\simeq 0.5$ for $ka \gtrsim 1$, in agreement with the statistical analysis of Section~\ref{sec:statistics}. Moreover, for $ka\lesssim1.4$, type-II vortices are considerably more likely to occur around a cylinder than conventional type-I vortices are to occur within a circle of the same radius in a random interference field.

Thus, laboratory measurements confirm our theoretical statistical model and demonstrate that subwavelength islands strongly promote the formation of wave vortices compared with homogeneous random wavefields.

\section{Conclusions}

To summarize, we have developed a statistical theory of type-II wave vortices around islands in random 2D wavefields, with or without the Coriolis effect, and verified its predictions by laboratory water-wave experiments. Unlike conventional (type-I) vortices, which arise at phase singularities in homogeneous interference wavefields, type-II vortices originate from wave scattering by subwavelength obstacles and can occur with remarkably high probabilities. 
In non-rotating random wavefields, the probability of a singly-charged type-II vortex approaches 50\%. The Coriolis effect further enhances vortex formation, producing resonant quasi-trapped modes, where the vortex probability approaches 100\%.

Strikingly, our simple statistical theory provides an explanation for the occurrence of M2-tidal vortices around major ocean islands (Iceland, Svalbard, Madagascar, and New Zealand), demonstrating that these oceanographic phenomena can be understood within a universal wave-scattering framework. 

Beyond ocean tides, our results reveal type-II vortices as a generic feature of random wavefields in multiply connected geometries. They suggest new opportunities for controlling wave topology and concentrating energy and OAM around subwavelength defects in a wide range of 2D wave systems, including plasmonic, acoustic, and electronic waves.

\begin{acknowledgements}
{\bf Funding:} K.B. was co-funded by the European Union through the project HORIZON-MSCA-2022-COFUND-01-SmartBRAIN3-101126600. L.S. is supported by National Key R\&D Program of China (2022YFA1404800 and 2023YFA1406900); National Natural Science Foundation of China (No. 12321161645, No. 12234007, No. 12221004, No. T2394480, T2394481); Science and Technology Commission of Shanghai Municipality (23DZ2260100).
{\bf Competing interests:} The authors declare no competing interests.
\end{acknowledgements}

\bibliography{refs}

\end{document}